\documentclass[12pt]{article}
\newcommand{\be}{\begin{equation}}
\newcommand{\ee}{\end{equation}}
\newcommand{\bea}{\begin{eqnarray}}
\newcommand{\eea}{\end{eqnarray}}
\newcommand{\ba}{\begin{array}}
\newcommand{\ea}{\end{array}}

\def\bbox{{\,\lower0.9pt\vbox{\hrule \hbox{\vrule height 0.2 cm
\hskip 0.2 cm \vrule height 0.2 cm}\hrule}\,}}
\newcommand{\dsl}{\pa \kern-0.5em /}

\newcommand{\bg}{\bar g}
\newcommand{\bR}{\bar R}
\newcommand{\bnabla}{\bar \nabla}

\begin{document}


\begin{titlepage}
\vfill
\begin{flushright}
\end{flushright}

\vfill
\begin{center}
\baselineskip=16pt
{\Large\bf Conserved gravitational charges from Yano tensors}
\vskip 1.0cm
{\large {\sl }}
\vskip 10.mm
{\bf David Kastor and Jennie Traschen} \\
\vskip 1cm
{

       Department of Physics\\
       University of Massachusetts\\
       Amherst, MA 01003\\
}
\vspace{6pt}
\end{center}
\vskip 0.5in
\par
\begin{center}
{\bf ABSTRACT}
\end{center}
\begin{quote}
The defining properties of Yano tensors naturally generalize those of Killing vectors to anti-symmetric tensor fields of arbitrary rank.  We show how the  Yano tensors of flat spacetime can be used to construct new, conserved gravitational charges  for transverse asymptotically flat spacetimes.  The relationship of these new charges to Yano tensors parallels that of ordinary ADM conserved charges to Killing vectors.  Hence, we call them Y-ADM charges.  A rank $n$ Y-ADM charge is given by an integral over a co-dimension $n$ slice of spatial infinity.  In particular, a rank $(p+1)$ Y-ADM charge in a $p$-brane spacetime is given by an integral over only the sphere $S^{D-(p+2)}$ surrounding the brane and may be regarded as an intensive property of the brane.
\vfill
\vskip 2.mm
\end{quote}
\end{titlepage}
\section{Introduction}

In this paper we introduce a new set of conserved gravitational charges for transverse asymptotically flat spacetimes, {\it e.g.} $p$-brane spacetimes that approach flat spacetime only  in directions transverse to the brane.  The usual ADM conserved charges \cite{Arnowitt:2004hi} of asymptotically flat spacetimes are closely associated with the translational and rotational Killing vectors of flat spacetime.  The new charges share this same association with Yano tensors \cite{yano},  anti-symmetric tensor fields that satisfy a natural generalization of Killing's equation.  For this reason, we call these new charges Y-ADM charges.  

An ADM charge is defined by an integral over a co-dimension $1$ slice of spatial infinity.  We will see that for a  Yano tensor of rank $n$, the corresponding Y-ADM charge is obtained by integrating only over a codimension $n$ slice of spatial infinity.   For  example, consider a  black $p$-brane in $D$ spacetime dimensions.   Further suppose that spatial infinity has either the topology $R^{p}\times S^{D-(p+2)}$, if the spatial directions tangent to the brane are infinite in extent, or  {\it e.g.} $T^p\times S^{D-(p+2)}$ if all of these directions are periodically identified on a torus.  The ADM mass gives the total mass of the spacetime, infinite in the first case and proportional to the volume of the torus in the second case.   In contrast, the Y-ADM charge, for the rank $(p+1)$-Yano tensor aligned with the  brane, is given by an integral only over the sphere $S^{D-(p+2)}$ surrounding the brane.  It is then on a similar footing to the electric charge of a $p$-brane, with respect to a $(p+1)$-form gauge potential, which is similarly given by an integral over only the $S^{D-(p+2)}$ surrounding the brane.  We see that while ADM charges measure extensive properties of a brane, Y-ADM charges can measure intensive quantities.  

For this reason, Y-ADM charges may be useful in formulating Bogomoln'yi bounds for branes.  Gibbons and Hull derived a  Bogomoln'yi bound for charged black holes \cite{Gibbons:fy} in $D=4$  by extending Witten's spinorial proof of the positive energy theorem \cite{Witten:mf} to include a Maxwell field.  This bound is saturated by $|Q|=M$ Reissner-Nordstrom black holes, which they showed have a super-covariantly constant spinor field.  A Bogomoln'yi bound for black brane spacetimes should similarly bound the mass of a black brane by its electric charge in some manner.  We usually infer the saturation of such bounds based on the existence of super-covariantly spinors.  However,  no explicit constructions of Bogomoln'yi bounds for black brane spacetimes have appeared in the literature.  

A priori, such bounds might be formulated in two different ways.  They could be stated in terms of extensive quantities, in which case the intensive electric charge would  be further integrated along the brane in order to be compared with the extensive ADM mass.  Alternatively, they might be stated in terms of intensive quantities, relating the electric charge to a Y-ADM gravitational charge, both given by integrals only over the  $S^{D-(p+2)}$ surrounding the brane.

Abbott and Deser (AD) \cite{Abbott:1981ff} showed that ADM charges may be constructed for any class of spacetimes that admit Killing vectors asymptotically.  The constuction of Y-ADM charges  presented below closely parallels the AD construction.  In the present paper we focus solely on Y-ADM charges for asymptotically flat spacetimes.  However, we expect that the construction may be extended to give Y-ADM charges for any spacetime that asymptotically admits Yano tensors.   The existence of Yano tensors in deSitter, anti-deSitter and reduced holonomy manifolds has been discussed recently in reference \cite{Cariglia:2003kf}.  Therefore, we expect Y-ADM charges may be defined for  asymptotically deSitter and anti-deSitter spacetimes, as well as wrapped brane spacetimes asymptotic to the product of flat spacetime with a reduced holonomy space.

Finally, we note that Yano tensors play an important role in the integrability of the Dirac equation in the Kerr-Newman spacetime \cite{Carter:fe} and related discussions of the supersymmetries of spinning point particles (see references \cite{Gibbons:ap} and \cite{Cariglia:2003kf} for extensive discussions and further references on these topics).

\section{Yano tensors}\label{counting}

The properties of Yano tensors \cite{yano}  generalize those of Killing vectors to rank $n$ antisymmetric tensor fields, with Killing vectors corresponding to the special case of rank $n=1$.  An $n$-form  $f_{a_1\dots a_n}=f_{[a_1\dots a_n]}$ is a Yano tensor \cite{yano} if it satisfies the equation $\nabla_{(a_1} f_{a_2)a_3\dots a_{n+1}}=0$, which implies the useful properties 
\bea
\nabla_{a_1} f_{a_2\dots a_{n+1}}&=&\nabla_{[a_1} f_{a_2\dots a_{n+1}]}\label{antisym}\\
\nabla_{a_1}f^{a_1\dots a_n}&=&0.\label{diverge}
\eea
It follows that second derivatives of Yano tensors are given by
\be\label{derivative}
\nabla_a\nabla_b f_{c_1\dots c_n}=(-1)^{n+1}{n+1\over 2} R^d_{\ a[bc_1}f_{c_2\dots c_n]d}
\ee
which reduces to the familiar formula for Killing vectors for $n=1$.

The maximum number of Yano tensors of rank $n$ in $D$ spacetime dimensions may be found by a straightforward extension of the argument  for determining the maximum number of Killing vectors (see {\it e.g.} appendix C of reference \cite{Wald:rg}).  As a consequence of equation (\ref{derivative}) a  Yano tensor is entirely determined by the value of its components and their first derivatives at any point.  If we start with $n=2$ for simplicity, then there are ${1\over 2}D(D-1)$ components of the Yano tensor $f_{ab}$ at a given point and ${1\over 6}D(D-1)(D-2)$ components of the first derivative $\nabla_a f_{bc}$ which must be totally antisymmetric.  Thus the maximum number of rank $2$ Yano tensors is then $N_2= {1\over 6}(D^3-D)$.  It is then also straightforward to see that the maximum number of Yano tensors of rank $n$ is given by
\be 
N_n = \left(  \begin{array}{c} D\\ n \end{array} \right) +\left(  \begin{array}{c} D\\ n+1 \end{array} \right).
\ee

Flat spacetime has the maximum number of Yano tensors of each rank.  Taking ranks $n=2$ as an example again, there are ${1\over 2}D(D-1)$ translationally invariant Yano tensors, which can be written as the $2$-forms
\be
f^{(ab)}=  dx^a\wedge dx^b
\ee
and ${1\over 6}D(D-1)(D-2)$ Yano tensors with components that are linear in the coordinates
\be
f^{(abc)}=  x^a dx^b\wedge dx^c + x^b dx^c\wedge dx^a + x^c dx^a\wedge dx^b. 
\ee
The Yano tensors $f^{(ab)}$, which we will call translational Yano tensors, are simply the wedge product of two translational Killing vectors. The Yano tensors $f^{(abc)}$, which we will call rotational Yano tensors, are built by taking the wedge product of a rotation, or boost,  in one plane with a translation in an orthogonal direction and then antisymmetrizing.  A similar construction for each rank yields the maximal number of Yano tensors in flat spacetime.

\section{AD construction for Yano tensors}\label{adconstruction}

The ADM mass, momenta and angular momenta of asymptotically flat spacetimes are associated respectively with the translational and rotational Killing vectors of flat spacetime.  Abbott and Deser  \cite{Abbott:1981ff} showed how to construct similar conserved charges for any spacetime that asymptotically admits Killing vectors.  For example, the AD construction yields conserved charges for asymptotically deSitter or anti-deSitter spacetimes.  

In this section we will see that the steps of the AD construction can be generalized to yield conserved charges for spacetimes that asymptotically admit Yano tensors.  Although we will eventually limit our scope to asymptotically flat spacetimes, we will begin working without this assumption.  The restriction to asymptotically flat spacetimes simplifies the construction considerably.  However, we expect that the construction can be extended to all spacetimes that asymptotically admit Yano tensors.  To simplify the presentation, in this section we consider only rank $2$  Yano tensors $f_{ab}$.  The construction of conserved charges for arbitrary rank Yano tensors is given in an appendix.  

\subsection{The conserved Yano current}\label{currentsection}
The AD construction begins by noting a property of spacetimes that exactly, as opposed to only asymptotically, admit Killing vectors.  If $f^a$ is a Killing vector, then the current given by
\be\label{killingcurrent}
j^a=G^{ab}f_b,
\ee
 is covariantly conserved.  This is easily seen by writing
\be\nabla_a j^a = (\nabla_a G^{ab}) f_b + G^{ab}\nabla_a f_b\label{conserve1}=0,
\ee
where the first term vanishes due to the well known divergence-less property  of  the Einstein tensor, and the second term vanishes using the symmetry of $G^{ab}$ and Killing's equation.  

Now consider instead a spacetime that exactly admits a rank $2$ Yano tensor $f_{ab}$.  It then follows that the two index anti-symmetric tensor 
\be\label{yanocurrent}
j^{ab}=-{1\over 4}\left(f^{cd}R_{cd}^{\ \ ab} - 2f^{ac}R_c^{\ b} + 2f^{bc}R_c^{\ a} + f^{ab} R\right )
\ee
is divergenceless, {\it i.e.} it satisfies
\be\label{divergecurrent}
\nabla_a j^{ab}=0.
\ee
Computing the divergence of $j^{ab}$ naturally yields a sum of eight terms.  Each of the four terms in which the derivative operator acts on the Yano tensor turn out to vanish due to the properties (\ref{antisym}) and (\ref{diverge}) of Yano tensors combined with the standard properties of the curvature tensors.  The four terms in which the derivative operator acts on a curvature tensor cancel after additionally making use of the Bianchi  identities\footnote{The first, uncontracted Bianchi identity is required only in the case of case of rank $n>2$ Yano tensors presented in the appendix.  }
\be\label{bianchi}
\nabla_{[a}R_{bc]de}=0, \qquad \nabla_a R_{bcd}^{\ \ \ a} + 2 \nabla_{[b} R_{c]d}\label{1contract}= 0,\qquad 
\nabla_a R^a_{\  b} - {1\over 2}\nabla_b R=0.
\ee
If instead a spacetime admits a rank $n$ Yano tensor, then a divergenceless, $n$ index anti-symmetric tensor $j^{a_1\dots a_n}$ may  be constructed.  An expression for $j^{a_1\dots a_n}$ is given in the appendix  in equation (\ref{bigcurrent}).  For the Killing vector case, $n=1$, the resulting conserved current is again that of equation (\ref{killingcurrent}) above.  We note that the overall numerical prefactor  in equation (\ref{yanocurrent}) for the Yano current is clearly arbitrary.  The factor $-1/4$ follows from formula (\ref{bigcurrent}) for the general case, where a particular rank $n$ dependent prefactor has been chosen to give agreement with the $n=1$ case.

Returning to the case of a rank $2$ Yano tensor, the conserved charge associated with the current $j^{ab}$ is most simply expressed in differential form notation.   If we consider the $2$-form 
\be
j={1\over 2}\,  j_{ab}\, dx^a\wedge dx^b,
\ee
then equation (\ref{diverge}) implies that $d{}^*j=0$.  If the total spacetime dimension is $D$, then the integral of  ${}^*j$ over any $(D-2)$-submanifold $\Sigma$ is conserved in the sense that it depends only on the homology class of $\Sigma$.

\subsection{The linearized Yano current in asymptotically flat spacetimes}
Thus far, we have considered a spacetime that  exactly admits a Yano tensor $f_{ab}$.  Following the AD construction, the next step is to consider a spacetime that admits a Yano tensor only asymptotically at spatial infinity.   Let $\bg_{ab}$ be a fixed background metric with covariant derivative operator $\bnabla_a$, having a rank $2$ Yano tensor $f_{ab}$.  We consider spacetimes asymptotic to $\bg_{ab}$ at infinity, with falloff at a rate sufficient to render the surface integral expressions for the charges defined below to be finite.  We write the full spacetime metric as 
\be
g_{ab}=\bg_{ab}+ h_{ab},
\ee
where $h_{ab}$ need not be small in the interior.
We will need the expression for the terms in the Riemann tensor linear in $h_{ab}$, which are given by
\bea
R^{L\ \ d}_{abc}&=&-2\bnabla_{[a}\Gamma^{L\ d}_{b]c}\\
\Gamma^{L\ d}_{bc}&=& {1\over 2}\left(\bnabla_b h_c^{\ d}+\bnabla_c h_b^{\ d}-\bnabla^d h_{bc}\right).
\eea
%
We will also need expressions for the linear terms in the Bianchi identities expanded around $\bar g_{ab}$.  At this point in the construction, we will specialize to the case of a flat background metric, {\it i.e.} $\bR_{abc}^{\ \ \ d}=0$.  With this assumption, the Bianchi identities (\ref{bianchi}) imply similar formulas for the linearized curvatures with respect to the background derivative operator
\be
\bnabla_{[a}R^L_{bc]de}=0,\qquad \bnabla_a R_{bcd}^{L \ \ a}+2 \bnabla_{[b} R^L_{c]d}= 0,\qquad 
\bnabla_a R^{La}_{\ \  b}- {1\over 2}\bnabla_b R^L=0.
\ee

Having specified that $g_{ab}=\eta_{ab}+h_{ab}$, the construction of Y-ADM charges applies to the  general class of  transverse asymptotically flat spacetimes.  In such spacetimes, the metric falls off to the flat metric as one approaches infinity in certain ``transverse" spatial directions, but not in other ``tangent" directions.   $P$-brane spacetimes, for example, approach the flat metric only in directions transverse to the brane.  Standard asymptotically flat boundary conditions can be regarded as a special case of transverse asymptotically flat boundary conditions.  

In the absence of an exact Yano tensor for the full metric $g_{ab}$, there is no Yano current $j^{ab}$ conserved with respect to the full covariant derivative operator.  However, following Abbott \& Deser, we consider the new current $k^{ab}$ defined in terms of the linearized curvatures by
\be\label{linearcurrent}
k^{ab}=-{1\over 4}\left(f^{cd}R_{cd}^{L\ ab} - 2f^{ac}R_{\ c}^{L\ b} + 2f^{bc}R_{\ c}^{L\ a} + f^{ab} R^L\right)
\ee
This new current $k^{ab}$ then satisfies the conservation law 
\be\label{nodiverge}
\bnabla_a k^{ab}=0
\ee
with respect to the background derivative operator and can be used to define a conserved charge.  
After plugging in the linearized curvature, we obtain a compact formula for the linearized current
\be\label{compact}
k^{ab}=3\, \delta^{abcd}_{klmn}\, f^{kl} \bnabla_c\bnabla^m h_d^{\ n},
\ee
where the symbol
$ \delta^{abcd}_{klmn}=
\delta^{[a}_{k}\delta^b_l\delta^c_m \delta^{d]}_{n}$
is totally anti-symmetric in both its up and its down indices.

We can now construct a conserved  charge from the $2$-form  $k= (1/2)k_{ab}\,dx^a\wedge dx^b$.
Equation (\ref{nodiverge}) implies that $d\, {}^{\bar\star} k=0$, where the symbol ${}^{\bar\star}$ denotes the Hodge dual with respect to the background metric.  The Y-ADM charge defined up to an arbitrary normalization factor by the integral of the $(D-2)$-form ${}^{\bar\star} k$ over a $(D-2)$ dimensional volume $\Sigma$, 
\be
 \int_\Sigma\  {}^{\bar\star} k,
\ee
is then conserved, {\it i.e.} invariant under changes of the volume $\Sigma$ that preserve its homology class.

The final step is to rewrite the Y-ADM charge as an integral over the $(D-3)$ dimensional boundary $\partial\Sigma$.  Since ${}^{\bar\star} k$ is closed, we can locally write 
${}^{\bar\star} k= d\,  {}^{\bar\star} l$ with $l$ a $3$-form.   The Y-ADM charge is then given, again up to a normalization factor to be specified below, by the boundary integral
\be\label{boundaryint}
\int_{\partial\Sigma }{}^{\bar\star} l.
\ee
We can find an explicit expression for $l=(1/6) l_{abc}dx^a\wedge dx^b\wedge dx^c$ in the following way.
The $2$-form current $k$ is related to the $3$-form $l$ by
$k^{ab}=\bnabla_a l^{abc}$.
Equation (\ref{compact})  can be straightforwardly rewritten as 
\be
k^{ab}=3\, \delta^{abcd}_{klmn}\, 
\left(  \bnabla_c(f^{kl}\bnabla^m h_d^{\ n}) -\bnabla^m((\bnabla_cf^{kl})h_d^{\ n}) 
+(\bnabla^m\bnabla_c f^{kl})h_d^{\ n}\right ).
\ee
In this expression, the third term within the parenthesis on the right hand side vanishes by virtue of equation (\ref{derivative}) and the assumption of a flat background.  The first term is manifestly the divergence of an anti-symmetric tensor, once the constant tensor $\delta^{abcd}_{klmn}$ is brought inside the parenthesis.  The second term is not obviously of this form.  However, if one expands the symbol $\delta^{abcd}_{klmn}$, many terms drop out due to the properties of Yano tensors, and the remaining terms have the desired form.  We then find that $k^{ab}=\bnabla_a l^{abc}$,  with
\be\label{boundary-linear}
l^{abc}= 3\, \delta^{abcd}_{klmn}\, f^{kl}(\bnabla^m h_d^{\ n}) 
-{1\over 4} (\bnabla^{[a} f^{bc]})h^d_{\ d}
+{3\over 4}(\bnabla^d f^{[ab})h^{c]} _{\ d}.
\ee
Note that for translational Yano tensors $\bar\nabla_c f_{ab}=0$ and only the first term contributes.

\subsection{Y-ADM charges}
As discussed above equation (\ref{boundaryint}), the Y-ADM charge associated with the Yano tensor $f^{ab}$ is obtained by integrating $l^{abc}$ over a codimension $2$ slice $\partial\Sigma$ of spatial infinity.  Given the relation $k^{ab}=\bnabla_a l^{abc}$, the charge can be written equivalently as an integral of $k^{ab}$ over $\Sigma$.  Let us write the background metric in the form 
\be
\bar g_{ab}= (n^c n_c)n_a n_b + x_a x_b + q_{ab},
\ee
where $n^a$ and $x^a$ are mutually orthogonal unit normal vectors to $\Sigma$, with $n^a$ allowed to be either timelike or spacelike, and $q_{ab}$ is the metric on $\Sigma$.  Let us further write near spatial infinity 
\be
q_{ab}= r_a r_b + \gamma_{ab}
\ee
where at spatial infinity $r^a$ and $\gamma_{ab}$ are the unit normal vector and metric on $\partial\Sigma$.
The Y-ADM charge $Y[f_{ab},\partial\Sigma]$ associated with the Yano tensor $f^{ab}$ and the surface of integration $\partial\Sigma$ at spatial infinity is then given in its boundary and  volume integral forms by
\bea
Y[f_{ab},\partial\Sigma]& \equiv & {1\over 8\pi}\int_{\partial\Sigma} d^{D-3}x  \sqrt{|\gamma|}\  n_{[a}x_{b}r_{c]}\  l^{abc}\\
&= &{1\over 8\pi} \int_\Sigma d^{D-2}x \sqrt{|q|}\  n_{[a}x_{b]}\ k^{ab} .
\eea
This equality is the main result of the paper.
The boundary integral expression is then the definition of the Y-ADM charge, applying both to spacetimes with smooth interiors and also to black brane spacetimes.   Equivalence of the boundary integral expression with the volume integral expression holds only in the case of smooth interiors, {\it i.e.} when there are no interior boundaries.
Similar formulas may be written for the Y-ADM charges corresponding to arbitrary rank Yano tensors.

Let us now establish some notation for a specific type of Y-ADM charges.  As described in section (\ref{counting}), the Yano tensors of flat spacetime can be written as linear combinations of basis elements of two types, translational and rotational.  Translational Yano tensors of rank $n$ are simply wedge products of $n$ translational Killing vectors.  Rotational Yano tensors are antisymmetrized combinations of  wedge products of a rotational Killing vector with $(n-1)$ translations.   Recall that to fully specify a Y-ADM charge, in addition to specifying the asymptotic Yano tensor, we must also choose a surface of integration at infinity.  Differently oriented surfaces of integration can yield different charges.
Let $Q^{(a_1\dots a_n)}$ denote the Y-ADM charge associated with the translational Yano tensor 
\be
f={1\over n!}\ dx^{a_1}\wedge\dots\wedge dx^{a_n},  
\ee
with the surface of integration taken to be orthogonal to each of the directions $x^{a_1},\dots,x^{a_n}$.  Consider the Killing vector case $n=1$.  The charge $Q^{(0)}$ is then the ADM mass, with a surface of integration that is a constant time slice at infinity.  The charge $Q^{(1)}$  is the ADM tension in the $x^1$ direction recently defined in references \cite{Traschen:2001pb}\cite{Townsend:2001rg} (see also \cite{Harmark:2004ch}).  In this case, the surface of integration includes the time direction, but excludes the $x^1$ direction.  Note that $Q^{(1)}$ defined in this way is distinct from the component of the ADM momentum in the $x^1$ direction, which is based on the same current, but is defined in terms of  integration over a constant time slice at infinity that includes the $x^1$ direction.

For spacetimes with smooth interiors, it is worth noting an important distinction between the volume integral expressions for Y-ADM charges in general and the special case of ADM charges, which correspond to rank $n=1$ Yano tensors.  For the case $n=1$ the volume integrand is precisely the linearized Einstein tensor contracted with the background Killing vector.  Using Einstein's equations this can be rewritten in terms of the stress energy tensor and the nonlinear terms in the Einstein tensor.  When these nonlinear terms are small, the ADM charges are given by integrals of components of the  stress energy tensor and acquire a straightforward physical interpretation in this way.   For   rank $n>1$ Yano tensors, we see from equation  (\ref{linearcurrent}) for rank $n=2$ and more generally from equation (\ref{generalcurrent}) that the volume integrand for the Y-ADM charge will include a contribution from the linearised uncontracted Riemann tensor.  This contribution cannot be rewritten in terms of the stress energy tensor and inhibits an equally  simple physical interpretation of the higher rank Y-ADM charges.  

In order to illustrate these comments, we give an explicit expression for the volume integral form of the charge $Q^{(01)}$,
\be
Q^{(01)}={1\over 16\pi}\int_\Sigma d^{D-2} x\sqrt{-\bar g} \left( (n^an^b-x^ax^b)R^L_{ab} -{1\over 2}R^L
-n^ax^bn^cx^d R^L_{abcd}
\right),
\ee
%
Using Einstein's equations, the terms in the integrand involving the Ricci tensor and scalar curvature can be rewritten in terms of components of the stress energy tensor and given a straightforward physical interpretation.  The final term, involving a particular component of the Riemann tensor cannot.    It would be quite interesting to find a physical interpretation of this latter contribution.

\section{Intensive vs. extensive charges}

In order to compare the results for ADM and Y-ADM charges, we consider the example of a static string in $D=5$ taken to lie along the $x^1$ axis.  We calculate the ADM mass $M=Q^{(0)}$, ADM tension $T=Q^{(1)}$ and also the Y-ADM charge $Q^{(01)}$ associated with the rank $2$ Yano tensor aligned with the string, {\it i.e.} with non-zero components $f^{01}=-f^{10}=1$.   The boundary  integral expressions for these charges, require only the metric far from the string.  This can be determined from  the linearized response to delta function sources extended along the $x^1$ axis and at the origin in the transverse $(x^2,x^3,x^4)$ space.  The nonzero components of the stress energy are taken to be
\be
T_{00}=\rho\, \delta^3(\vec x),\qquad T_{11}=\lambda\, \delta^3(\vec x)
\ee
where $\vec x$ denotes position in the transverse $3$-space.  

This example will  illustrate the intensive nature of Y-ADM charges.  The charges we calculate will, of course, all be proportional to the parameters $\rho$ and $\lambda$ that characterize the string source.  However,  they differ both in the details of the boundary terms and in the dimensions of the surfaces of integration. 
Spatial infinity, in this example, has topology $R^2\times S^2$, where the $R^2$ directions are the $(x^0,x^1)$ directions tangent to the worldvolume of the string.  Calculating the ADM mass  requires an integral over the full spatial cylinder $R\times S^2$  surrounding the  string at infinity.  It is thus an extensive property of the brane.  If  the $x^1$ direction is periodically identified with length $\Delta L$, then the ADM mass will be proportional to $\rho\, \Delta L$.   The parameter $\rho$ then is then proportional to a mass per unit length.  

The ADM tension, which also  requires an integral over a surface with topology $R\times S^2$, is similarly an extensive quantity.  The extended direction, in this case however,  is now the time direction.  Let the integral to be restricted to a finite time interval $\Delta t$.  The tension will then be proportional to $\lambda\, \Delta t$ and the parameter $\lambda$ is proportional to a tension per unit time.  In contrast, calculating the Y-ADM charge $Q^{(01)}$ requires only an integral over a spatial $S^2$ at constant $x^0$, $x^1$.  It will simply be a linear combination of the parameters $\rho$ and $\lambda$.  These points are illustrated by the following calculations.

The non-trivial components of the linearized Einstein equations with the sources specified above are given by
\bea
\partial^k\partial_k \bar h_{00} &=& -16\pi\, \rho\, \delta^3(\vec x)\\
\partial^k\partial_k \bar h_{11} &=& -16\pi\, \lambda\, \delta^3(\vec x)
\eea
where $\bar h_{ab}=h_{ab}-1/2\eta_{ab}h$.    The solution for a static string then has the nonzero components
\be 
h_{00}= {4(2\rho+\lambda)\over 3r},\qquad h_{11}= {4(\rho+2\lambda)\over 3r},\qquad h_{kl}= {4(\rho-\lambda)\over 3r}\delta_{kl}
\ee
where $k,l=2,3,4$.  We note two special sets of values for the parameters $\rho$ and $\lambda$.  Setting $\lambda=-\rho$ gives $h_{00}=-h_{11}$, which is a boost invariant string.   Setting $\lambda=-\rho/2$ gives $h_{11}=0$ and $h_{00}=h_{kk}$ which is the asymptotic form of the uncharged black string obtained by crossing $D=4$ Schwarzschild  with a line.

Let us first calculate the Y-ADM charge $Q^{(01)}$ for the Yano tensor aligned with the string, which is given after plugging in the appropriate normal vectors  by
\be
Q^{(01)}= {1\over 8\pi}\int d^2\Omega \ {x^k\over r}\  l^{01k},
\ee
where $r^2=(x^2)^2+(x^3)^2+(x^4)^2$.  
From equation (\ref{boundary-linear})  we find that
\be
 l^{01k}=-{1\over 4}( \partial^l h_l^{\ k}-\partial^k h_l^{\ l})
\label{boundaryterm}
\ee
giving the result
$Q^{(01)}= - (\rho-\lambda)/3$.
The ADM mass of the string is given by
\be
M= {1\over 8\pi}\int dx^1 d^2\Omega\  {x^k\over r}\  l^{0k}.
\ee
The quantity $l^{0k}$ can be found from equation  (\ref{generalboundary})
\be
l^{0k}  = {1\over 2}(\partial^\alpha h_\alpha^{\ k}-\partial^k h_\alpha^{\ \alpha} ),
\ee
where the index $\alpha$ runs over the values $(1,2,3,4)$.  The result of the integration gives
$M =  \rho \Delta L$ where $\Delta L$ is the length of the $x^1$ direction.  The ADM tension is given by
\be
T= {1\over 8\pi}\int dx^0 d^2\Omega {x^k\over r} l^{1k}.
\ee
where again from equation (\ref{generalboundary}) we find
\be
l^{1k}={1\over 2}(\partial^\alpha h_\alpha^{\ k}-\partial^k h_\alpha^{\ \alpha})
\ee
where in this case the index $\alpha$ runs over the $(0,2,3,4)$.  The result of the integration is then 
$T=-\lambda\Delta t$ where $\Delta t$ is the length of the time integration.  This example illustrates that the ADM charges are extensive quantities, while the Y-ADM charge aligned with the brane is an intensive quantity.  

\section{Conclusions}

In this paper we have constructed a new set of gravitational charges based on Yano tensors.   
We saw that the properties of Yano tensors allow for a simple extension of the Abbott \& Deser method of constructing ADM charges \cite{Abbott:1981ff}.  Although we have focused on transverse asymptotically flat spacetimes in this paper, we expect that Y-ADM charges may be constructed for any class of spacetimes that asymptotically admit Yano tensors.  Such backgrounds would include deSitter, anti-deSitter and products of flat spacetime with reduced holonomy spaces, such as Calabi-Yau compactifications.   A second direction for future work is to ask whether Y-ADM charges arise as boundary terms in a suitable generalization of the Hamiltonian formulation of general relativity.  In such a formulation, one could consider the evolution of data on $D-n$ dimensional slices under an $n$-parameter flow.

One interesting  property of  certain Y-ADM charges for  $p$-brane spacetimes, which we have highlighted,  is that they are intensive quantities, {\it i.e.} not proportional to the volume of the brane like the ADM mass.  In this sense, they are analogous to the electric or magnetic charge of a $p$-brane.  For this reason, we have speculated that Y-ADM charges may arise in extending the formulation of BPS bounds for charged black holes \cite{Gibbons:fy} to charged black branes. 

We conclude with some observations that may prove relevant to this last topic.  Given the general interest in black brane spacetimes,  it is natural to ask whether these spacetimes themselves admit Yano tensors?  If we consider the case of Killing vectors, then the Schwarzshild family of spacetimes all have static Killing vectors.   Does a black $p$-brane spacetime similarly admit a rank $p+1$ Yano tensor exactly, and not just asymptotically? 

Consider the simple example of a black string spacetime in $D=5$ obtained by adding a flat direction to the $D=4$ Schwarzschild spacetime
\be
ds^2 = - (1-{2m\over r}) dt^2 + dy^2  + (1-{2m\over r})^{-1}dr^2 +r^2d\Omega_2^2.
\ee
One can check that this spacetime does not have a Yano tensor  of the form
\be 
f= F(r) dt\wedge dy.
\ee
Moreover, the detailed calculation makes clear that the only black brane spacetimes having such an aligned Yano tensor will be those with boost invariance in the directions tangent to the brane.  For example,  assume that the metric has the general form 
\be
ds^2= h(x^\rho)\, \eta_{ab}\, dx^adx^b+ k(x^\rho)\, \delta_{\alpha\beta}\, dx^\alpha dx^\beta
\ee
where $a,b=0,1,\dots,p$ are directions parallel to the brane, $\alpha,\beta=p+1,\dots,D-1$ are directions transverse to the brane, and the functions $h$ and $k$ depend only on the transverse coordinates.  It is now straightforward to show that 
\be
f= h^{(p+2)/ 2} dx^0\wedge dx^1\wedge\dots\wedge dx^{p}
\ee
is a Yano tensor.  It is interesting to note that, except in the case $p=0$, the Yano tensor $f$ it is not equal to the wedge product of the Killing vectors tangent to the brane.  The case $p=0$, {\it e.g.} Schwarzschild spacetime,  is a degenerate one, because there are no boosts to consider along directions tangent to a $0$-brane.   Since extremal branes are generally boost invariant, this result may be relevant for the actual construction of BPS bounds for $p$-brane spacetimes.  Note that in addition to extremal  black branes, the Melvin spacetime \cite{Melvin:1963qx} and its various fluxbrane generalizations provide another example of boost invariant spacetimes.

\vskip 1.0cm
\noindent{\bf Acknowledgement} This work was supported in part by National Science Foundation grant NSF PHY0244801.

\appendix

\renewcommand{\thesection}{Appendix \Alph{section}}

\section{Conserved charges for arbitrary rank Yano tensors}\label{app1}
In section (\ref{adconstruction}) above, we have given results only for rank $2$ Yano tensors.  Similar results hold for arbitrary rank Yano tensors and will be given in this appendix.
The conserved Yano current for a spacetime admitting a  rank $n$ Yano tensor $f_{a_1\dots a_n}$ is given by
\be\label{bigcurrent}
j^{a_1\dots a_n} =   -{(n-1)\over 4}  R^{[a_1a_2}_{\ \ \ \ \ bc}f^{a_3\dots a_n]bc}
+(-1)^{n+1}R_c^{\ [a_1}f^{a_2\dots a_n]c}
- {1\over 2n} R f^{a_1\dots a_n}, 
\ee
where the overall normalization of the current has been set to agree with equation (\ref{killingcurrent}) in the case $n=1$.  The Yano current can also be written in the more compact form
\be
j^{a_1\dots a_n} =N_n\,  \delta^{a_1\dots a_n kl}_{b_1\dots b_n pq} \, 
f^{b_1\dots b_p} R_{kl}^{\ \ \ pq},  
\ee
where the symbol $\delta^{a_1\dots a_m}_{b_1\dots b_m}$ is defined to be
\be\label{antisym-tensor}
\delta^{a_1\dots a_m}_{b_1\dots b_m}=
\delta^{[a_1}_{b_1}\cdots \delta^{a_m]}_{b_m}
\ee
is totally anti-symmetric in both its up and its down indices and the constant $N_p$ is given by
\be
N_n=-{(n+1)(n+2)\over 4n}.
\ee
The linearized current for a transverse asymptotically flat spacetime is given by
\be\label{generalcurrent}
k^{a_1\dots a_n}= 
-2 N_n\,  \delta^{a_1\dots a_n kl}_{b_1\dots b_n pq} \, 
f^{b_1\dots b_p} \bar\nabla_k\bar\nabla^p h_l^{\ q} .
\ee
The linearized current may be written as a divergence of a totally antisymmetric tensor in the following way.  Start by rewriting equation (\ref{generalcurrent}) as
\be\label{generallinear}                                                                                                                                                                                                                                                                                                                                                                                                                                                                                                                       
k^{a_1\dots a_n} = 
-2N_n \delta^{a_1\dots a_n kl}_{b_1\dots b_n pq} \left(
\bar\nabla_k(f^{b_1\dots b_n} \bar\nabla^p h_l^{\ q} )
-\bar\nabla^p(h_l^{\ q}\bar\nabla_k f^{b_1\dots b_n})\right )
\ee
We would like to write this in the form $k^{a_1\dots a_n}=\nabla_b l^{ba_1\dots a_n}$ with 
$l^{ba_1\dots a_n}=l^{[ba_1\dots a_n]}$.
The first term inside the parenthesis in equation (\ref{generallinear}) is manifestly the divergence of an antisymmetric tensor.  The second term is not manifestly so, but as in the $n=2$ case reduces to such an expression after expanding the symbol (\ref{antisym-tensor}) and making use of the general properties of Yano tensors.  The final result is
\be\label{generalboundary}
l^{k a_1\dots a_n}= -2N_n \delta^{a_1\dots a_n kl}_{b_1\dots b_n pq}f^{b_1\dots b_n} \bar\nabla^p h_l^{\ q} -{1\over 2n}\left( h_l^l \bar\nabla^k f^{a_1\dots a_n}-(n+1)h^{l[k}\bar\nabla_l f^{a_1\dots a_n]}\right).
\ee
Note that the case $n=1$ gives a useful, covariant form for the ADM boundary term.  Explicit expressions for the corresponding Y-ADM charges can simply be written down by extending the formulas for the case of rank $n=2$ given in the main text.


\begin{thebibliography}{99}

\bibitem{Arnowitt:2004hi}
R.~Arnowitt, S.~Deser and C.~W.~Misner,
``The Dynamics of General Relativity,''  in ``Gravitation: an introduction to current research,"  Louis Witten  ed., Wiley (1962) 
[arXiv:gr-qc/0405109].


\bibitem{yano}
K.~Yano, ``Some Remarks on Tensor Fields and Curvature," Ann.\ Math.\ {\bf 55}, 328 (1952).


\bibitem{Gibbons:fy}
G.~W.~Gibbons and C.~M.~Hull,
``A Bogomolny Bound For General Relativity And Solitons In N=2 Supergravity,''
Phys.\ Lett.\ B {\bf 109}, 190 (1982).

\bibitem{Witten:mf}
E.~Witten,
``A Simple Proof Of The Positive Energy Theorem,''
Commun.\ Math.\ Phys.\  {\bf 80}, 381 (1981).

\bibitem{Abbott:1981ff}
L.~F.~Abbott and S.~Deser,
``Stability Of Gravity With A Cosmological Constant,''
Nucl.\ Phys.\ B {\bf 195}, 76 (1982).

\bibitem{Cariglia:2003kf}
M.~Cariglia,
``Quantum mechanics of Yano tensors: Dirac equation in curved spacetime,''
Class.\ Quant.\ Grav.\  {\bf 21}, 1051 (2004)
[arXiv:hep-th/0305153].

\bibitem{Carter:fe}
B.~Carter and R.~G.~Mclenaghan,
``Generalized Total Angular Momentum Operator For The Dirac Equation In Curved
Space-Time,''
Phys.\ Rev.\ D {\bf 19}, 1093 (1979).

\bibitem{Gibbons:ap}
G.~W.~Gibbons, R.~H.~Rietdijk and J.~W.~van Holten,
``Susy In The Sky,''
Nucl.\ Phys.\ B {\bf 404}, 42 (1993)
[arXiv:hep-th/9303112].

\bibitem{Wald:rg}
R.~M.~Wald,
``General Relativity,''
U. Chicago Press (1984).

\bibitem{Traschen:2001pb}
J.~H.~Traschen and D.~Fox,
``Tension perturbations of black brane spacetimes,''
Class.\ Quant.\ Grav.\  {\bf 21}, 289 (2004)
[arXiv:gr-qc/0103106].

\bibitem{Townsend:2001rg}
P.~K.~Townsend and M.~Zamaklar,
``The first law of black brane mechanics,''
Class.\ Quant.\ Grav.\  {\bf 18}, 5269 (2001)
[arXiv:hep-th/0107228].

\bibitem{Harmark:2004ch}
T.~Harmark and N.~A.~Obers,
``General definition of gravitational tension,''
arXiv:hep-th/0403103.

\bibitem{Melvin:1963qx}
M.~A.~Melvin,
``Pure Magnetic And Electric Geons,''
Phys.\ Lett.\  {\bf 8}, 65 (1964).

\end{thebibliography}
\end{document}